\documentstyle[12pt]{article}
\begin{document}
\renewcommand
\baselinestretch{1.5}
\textwidth=17cm

\begin{flushright}
FTUAM 95/24\\
June 1995
\end{flushright}
\vspace{2cm}
\begin{center}
{\LARGE Quantum Mechanical Lorentzian Wormholes\\  in
Cosmological Backgrounds}
\end{center}
\vspace{2cm}
\begin{center}
{\large David Hochberg\footnote{\large Email: hochberg@ccuam3.sdi.uam.es}}
\end{center}
\begin{center}
{\large \sl Departamento de F\'isica Te\'orica, Universidad
 Aut\'onoma de Madrid\\
Cantoblanco, 28049 Madrid, Spain}
\end{center}

\vspace{2cm}
\begin{abstract}
We present a minisuperspace analysis of a class of Lorentzian
wormholes that evolves quantum mechanically in a background
Friedman Robertson Walker spacetime. The quantum mechanical
wavefunction for these wormholes is obtained by solving the
Wheeler-DeWitt equation for Einstein gravity on this minisuperspace.
The time-dependent expectation value of the wormhole throat radius is
calculated to lowest order in an adiabatic expansion of the
Wheeler-DeWitt hamiltonian. For a radiation dominated expansion, the
radius is shown to relax asymptotically to obtain a value of order
the Planck length while for a deSitter background, the radius is
stationary but always larger than the Planck length. These two cases are
of particular relevance when considering wormholes in the early universe.

\end{abstract}
\vfill\eject

\section{Introduction}

   Wormholes are handles in the spacetime topology linking widely
separated regions of the universe, or ``bridges'' joining two or more
different spacetimes. Current interest in Lorentzian or
Minkowski-signature wormholes has been motivated initially by attempts to
understand nontrivial spacetime topologies, topology changing
processes, their physical consequences,
and their role, if any, in quantum gravity.
A survey of recent work on
wormhole physics includes topics
addressing the fundamental properties of traversable
macroscopic
wormholes \cite{Morris88a,Morris88b},
construction of explicit wormhole solutions and
corresponding classical \cite{Visser89}
and quantum \cite{Visser90a}  stability analysis,
wormholes as time machines
and causality violation \cite{Visser90b} , wormholes in
higher-derivative gravity \cite{Hoch90} , the phenomenon of gravitational
squeezing of the vacuum as a natural mechanism for producing the states of
weak energy condition violating matter required to prop open the
throats of semiclassical microscopic
wormholes \cite{Hoch91} , wormholes as
representing one possible mode of
fluctuation of the spacetime foam \cite{Redman94} , and
wormholes as gravitational lenses \cite{Cramer95}.

Recently, a very simple model of classical dynamic Minkowski signature
wormholes connecting Friedman-Robertson-Walker (FRW) spacetimes was
constructed with the aim of initiating an investigation of the
possible
cosmological effects of such objects in the early
universe \cite{Hoch93} . In
that
work, an alternative solution to the horizon problem was proposed
based on the possibility that the early universe did not necessarily
inflate, but was populated with a network of microscopic Planck scale
wormholes joining otherwise causally disjoint regions of space.
The essence of the idea being that these wormholes can allow two-way
transmission of
signals between spatially separated regions of spacetime and thus
permit
such regions to come into thermal contact.
This wormhole
``phenomenology'' is based on very elementary considerations. Indeed,
to be
of cosmological interest, these wormholes need only
have stayed open long enough
for the radiation, initially located in causally disconnected regions, to
traverse
the ``handle''.
Within the class of wormholes treated in \cite{Hoch93}, a crude
estimate for the time required for thermalization can be had
considering the
following purely illustrative example. First, assume an initial number
density $n(t_{Pl})$ of Planck-sized wormholes of average radius $L_P$
at the Planck time and let $n(t_{Pl}) = \gamma \, n_{Pl}$, where
$\gamma = {\rm const.}$ represents a fraction of the Planck number
density: $n_{Pl} = (L_P)^{-3}$. Then, at some later time $t$,
$n(t) \sim R^{-3}$, so
the wormhole
number density is given by
\begin{equation}
n(t) = \gamma \, n_{Pl} \, \left( \frac{R_{Pl}}{R(t)} \right)^3.
\end{equation}
Define the volume-filling factor (the volume of space filled by the
interior of wormhole mouths) by
\begin{equation}
v_{fill}(t) = \frac{{\rm number\, density} \times {\rm wormhole
\,volume}}{{\rm Planck\, density} \times {\rm Planck\, volume}} =
\frac{n(t)}{n_{Pl}}\, \left( \frac{l(t)}{L_P} \right)^3.
\end{equation}
The instantaneous wormhole throat radius $l(t) = aR$, where $R$ is the scale
factor of the background spacetime and $a$ is the wormhole radius in
the limit of a static background. For a radiation dominated cosmology
and with a matter distribution on the throat
subject to a particular equation of state
 ($P$ is the surface pressure density and $\sigma$ the surface energy
density on the throat)
$P = P(\sigma)$, one can show that the throat decouples from
the expansion with
$l \sim R^2$, so that
the filling factor becomes
\begin{equation}
v_{fill}(t) = \gamma \, \left( \frac{R(t)}{R_{Pl}} \right)^3.
\end{equation}
When $v_{fill}(T) \approx 1$, there is one wormhole per unit volume
by this equilibrium time $T$, and the average particle
will have traversed at least one
wormhole.
So, for example, an initial wormhole number density of only
$\gamma \approx 10^{-14}$ allows thermalization of distant spatial
regions by the GUT time scale $T_{GUT} \sim 10^{-34} {\rm sec}$.
Of course, different values of $\gamma$ will lead to different values
for the thermalization time scale $T$.

While the phenomenological application put
forward in \cite{Hoch93} is novel, it
is far from constituting a proof as such, and much more work is
clearly
required to settle the issue.
As a case in point, note that a constant $\gamma$ implies the total
number of wormholes is constant. This need not be the case in general
and we expect $\gamma = \gamma (t)$. This quantity is related to the
initial spectrum of wormholes and a first principles calculation of
this factor would provide significant insight into the structure of
spacetime at the Planck scale. Of equal importance is the fact that
the wormholes treated
in \cite{Hoch93}  were strictly classical and if such objects do indeed exist
 at or near the Planck scale, it is much
more likely they will be quantum mechanical in nature.
Moreover, one would like to know how the wormholes evolve in time and
how the specifics of their temporal development is affected by the
background spacetime.
It is this
latter issue which shall be taken up in the present paper.

The quantum mechanics of Minkowski signature wormholes in FRW
backgrounds will be addressed within the framework of a minisuperspace
analysis applied to a specific class of wormholes resulting from
surgically
modified FRW cosmologies. The wavefunction of the wormhole will be
obtained as a solution of the Wheeler-DeWitt equation.
Because no one knows how to solve the
Wheeler-DeWitt equation on the full extended superspace of three
dimensional
metrics (and matter field configurations), it proves extremely useful
to examine this equation on minisuperspace configurations, which are
restricted
configurations described by only a finite number of
degrees-of-freedom, with all the other modes of the full superspace
frozen
out. This is a standard approximation employed in quantum cosmology
and is
easily adapted for quantizing localized configurations in spacetime. Such
analyses have
been applied to the study of quantum wormholes in Minkowski,
Nordstrom-Reissner and Schwarzschild-deSitter
backgrounds \cite{Visser90a,Kim92}.
 While the Wheeler-DeWitt equation
is time-independent, the expanding background provides a natural time
parameter which an external observer can employ in order to
characterize change in the wormhole's configuration. For FRW
spacetimes, this time parameter is simply the scale factor (or any
invertible
function of the scale factor).

This paper is organized as follows. In Section II we describe in
detail the minisuperspace model we adopt for Lorentzian wormholes.
Canonical quantization of the wormhole is carried out in Section III
and the Wheeler-DeWitt equation solved in perturbation theory for the
wormhole wavefunction. The expectation value of the wormhole throat
radius is calculated and is found to be partially dragged along with the
expansion
when the universe is radiation dominated, although relaxing to attain
a value of the order the Planck length for late times. By contrast,
a quantum wormhole in a deSitter background maintains a constant
throat radius that is always greater than the Planck length. Both
these statements hold when there is no matter residing on the wormhole
throat.
We briefly comment on the effects of including matter in Section IV.
Discussion and caveats are presented in Section V. A Green function
needed
for the solution of a certain nonlocal differential equation is
calculated
in the Appendix. Units with $G = c = 1$ are used throughout, while
$\hbar = L^2_P$, where $L_P$ is the Planck length.

\section{Minisuperspace Model for Wormholes}

The wormholes we will be considering result from surgically modified
Friedman-Robertson-Walker spacetimes. We adopt this technique
to make the subsequent analysis tractable, however, we assume the qualitative
features of the wormholes are independent of the details of the
construction. To construct them, take two copies $M_1, M_2$ of a FRW
spacetime with identical \cite{foot1} scale factors $R(t)$
and spatial curvature
constant $\kappa$:
\begin{equation}
ds^2 = -dt^2 + R^2(t) \left[ \frac{dr^2}{1 - \kappa r^2} + r^2 (d\theta^2 +
  \sin^2 \theta \, d\phi^2) \right] ,
\end{equation}
and remove from each an identical four-dimensional region of the form
$\Omega_{1,2}= \{ r_{1,2} < a \}$. The resulting spacetime contains
two disjoint boundaries $\partial \Omega_{1,2} = \{ r_{1,2} = a \}$
which are timelike hypersurfaces. An orientation preserving
identification
$\partial \Omega_1 = \partial \Omega_2 = \partial M$ yields two FRW cosmologies
connected by a wormhole whose throat is located on their mutual
boundary $\partial \Omega$.
This model is an extreme version of a wormhole for which the spacetime
curvature in the throat is much greater than that in the regions
surrounding the mouths.
Here $a = a(\tau)$, is a function
describing the dynamics of the wormhole's throat.
It is important to point out that the proper motion of the wormhole
throat is in general independent, that is, decoupled, from the
background cosmic flow. Nevertheless, at any given instant, the
physical radius (see the explicit form of the throat's three-metric
below in (9))
of the wormhole is given by the product $aR$.
 The wormhole is spherically
symmetric and the boundary layer is just the world volume swept out by
its throat. This procedure also leads to a wormhole connecting a
single FRW spacetime to itself if one identifies the two background
geometries, i.e., we have a FRW spacetime with a handle. In this case,
the two regions $\Omega_{1,2}$ can be separated by an arbitrarily
large distance in an open $(\kappa =0, -1)$ universe.
For the classical case, one insures that the modified spacetime is
itself a solution of the gravitational field equations by effecting a
proper matching of the metric across the boundary layer \cite{Israel67} .
These classical wormholes must necessarily involve stress-energy
distributions that violate the weak energy condition in the vicinity
of the throat. This feature is, however, not an artifact of the
construction and holds in general \cite{Morris88a, Morris88b}.
At the
quantum mechanical level, the state of the wormhole is described by appropriate
solutions of the
Wheeler-DeWitt equation, either with, or without the inclusion of matter.

As a consequence of the above construction, the Ricci tensor is everywhere
FRW except at the wormhole throat:
\begin{equation}
R^{\mu}_{\nu}(x) = \left( \begin{array}{cc}
                   Tr \Delta {\cal K} & 0 \\
                   0  &  \Delta {\cal K}^i_j
                  \end{array}
                  \right)\delta(\eta)
 + R^{(1)\mu}_{\nu}(x) \, \Theta (\eta)
 + R^{(2)\mu}_{\nu}(x) \, \Theta (-\eta),
\end{equation}
where $R^{(1)\mu}_{\nu}, R^{(2)\mu}_{\nu}$ are the Ricci tensors for
the two spacetimes $M_1$ and $M_2$, respectively
and $\Delta {\cal K}^i_j = ({\cal
K}^{(2)i}_{j} - {\cal K}^{(1)i}_{j})$ is the jump, or discontinuity, in the
extrinsic
curvature of the boundary layer (i.e., the throat) in
transiting from $M_1$ to $M_2$.
Proper normal distance as measured from the throat is conveniently
parametrized in terms of the Gaussian normal coordinate $\eta$. For
the case at hand, reflection symmetry  implies $\Delta {\cal K}^i_j =
2 {\cal K}^{(2)i}_j$ and spherical symmetry implies that
${\cal K}^i_j = {\rm diag} ({\cal K}^{\tau}_{\tau}, {\cal
K}^{\theta}_{\theta}, {\cal K}^{\theta}_{\theta})$. The gravitational
action
for the wormhole is given by two contributions, to whit
\begin{equation}
S_{wormhole} = S_{throat} + S_{background},
\end{equation}
where
\begin{equation}
S_{throat} = -\frac{1}{8 \pi} \int_{\partial M} d^3x \, \sqrt{-h}\,
\, Tr(\Delta {\cal K}),
\end{equation}
and
\begin{equation}
S_{background} = -\frac{1}{16\pi}\int_{M_2} d^4x \, \sqrt{-g} \,R^{(2)}
 - \frac{1}{16\pi} \int_{M_1} d^4x \, \sqrt{-g} \,R^{(1)}.
\end{equation}

 We shall first calculate the throat action. The 3-dimensional metric
$h_{ij}$ induced on the throat world-volume is
\begin{equation}
h_{ij}(x) = {\rm diag} \left( -1, a^2(\tau)R^2(t), a^2(\tau)R^2(t)
 \sin^2 \theta \right),
\end{equation}
where $\tau$ denotes the proper time on the wormhole throat and $t$
refers to the background cosmic time. To evaluate (7), we need the
components of the extrinsic curvature tensor
\begin{equation}
{\cal K}_{ij} = n_{\mu} \nabla_{(j)} e^{\mu}_{(i)},
\end{equation}
where the ${\bf e}_{(i)}$ constitute a set of three linearly
independent
tangent vectors defined along the intrinsic coordinates $\xi^i$
parametrizing the world-volume hypersurface and $n^{\mu}$ is the outward unit
normal
$(n^{\mu} n_{\mu} = 1)$. The covariant derivative intrinsic to the
hypersurface is taken along the $jth$ coordinate direction. The
throat proper time and the two angles provide a convenient set of
intrinsic coordinates: $\xi^i = (\tau, \theta, \phi)$. As the location
of the throat (i.e., its embedding with respect to the background FRW
geometry) is $X^{\mu} = (t,a,\theta, \phi)$, the tangent vectors
are given by $e^{\mu}_{(i)} = dX^{\mu}/d{\xi}^i$. Since $n_{\mu}
e^{\mu}_{(i)} = 0$ for $i=1,2,3$, the unit normal to the throat is given by
\begin{equation}
n^{\mu} = \left ( \frac{{\dot a}R}{\sqrt{1 - \kappa a^2}},
 \frac{1}{R} \sqrt{1 - \kappa a^2 + {\dot a}^2 R^2}, 0, 0 \right ).
\end{equation}
A straightforward calculation of (10) yields
\begin{equation}
{\cal K}^{\phi}_{\phi} \,= \,
{\cal K}^{\theta}_{\theta} = \frac{{\dot a}R'}{\sqrt{1 - \kappa a^2}} +
 \frac{1}{aR} \sqrt{1 - \kappa a^2 + ({\dot a} R)^2},
\end{equation}
and
\begin{equation}
{\cal K}^{\tau}_{\tau} = \frac{{\ddot a}R}{\sqrt{1 -\kappa a^2 +({\dot
a}R)^2}}
 + \frac{2{\dot a}R'}{\sqrt{1 - \kappa a^2}}
 + \frac{(\kappa a){\dot a}^2 R}{(1 - \kappa a^2)
 \sqrt{1 - \kappa a^2 + ({\dot a} R)^2 }},
\end{equation}
where ${\dot a} = da/d\tau$ and $R' = dR/dt$. Note the appearance of
the two time parameters $\tau$ and $t$ in (12) and (13). We can always
cast the equations in terms of one or the other time parameter by
means of the jacobian
\begin{equation}
e^0_{\tau} = \frac{dt}{d\tau} = \left( \frac{1 - \kappa a^2 +
 ({\dot a} R)^2}{1 - \kappa a^2} \right )^{1/2}.
\end{equation}
Physically, this amounts to a calibration of the clocks attached to
the throat in terms of the comoving clocks, if we choose to express
$\tau$ in terms of $t$, or vice-versa, if instead we choose to express
$t$ as a function of $\tau$. We shall have occasion to use (14) below.
{}From (7), (9), (12) and (13), we have $Tr({\cal K}) = \Delta {\cal
K}^{\tau}_{\tau} + 2\, \Delta {\cal K}^{\theta}_{\theta}$ and
\begin{eqnarray}
S_{throat} = \int d\tau [ 2(a{\dot a}R^2  &+&  a^2 R {\dot R})
 \sinh^{-1} \left( \frac{{\dot a}R}{\sqrt{1 - \kappa a^2}} \right)
 - \frac{3{\dot a}R'(a^2 R^2)}{\sqrt{1 - \kappa a^2}}        \\
 &-&   2aR \sqrt{1 - \kappa a^2 + ({\dot a}R)^2}  ] ,  \nonumber
\end{eqnarray}
where we have made use of the identity
\begin{eqnarray}
\frac{{\ddot a}R}{\sqrt{1 - \kappa a^2 + ({\dot a}R)^2}} =
 \frac{d}{d\tau} \sinh^{-1} \left( \frac{{\dot a}R}{\sqrt{1 - \kappa
a^2}} \right) & - & \frac{{\dot a}{\dot R}}{\sqrt{1 - \kappa a^2 +
 ({\dot a}R)^2}} \nonumber \\
 & - & \frac{\kappa a{\dot a}^2 R}{(1 - \kappa a^2)\sqrt{1 - \kappa a^2
+({\dot a} R)^2}},
\end{eqnarray}
together with an integration by parts. Because ${\dot R} =
R'(dt/d\tau)$, we are able to express ${\dot R}$ in terms of $R,R', a$
and
${\dot a}$.

The contribution to the action coming from the background (external to
the throat) spacetime is
\begin{equation}
S_{background} = 3\int_{M}
 dt \,
 \frac{dr\, r^2}{\sqrt{1 -
\kappa
 r^2}} (R^2 R'' + R {R'}^2 + \kappa R),
\end{equation}
where
$M = (M_2 - \Omega_2) \bigcup (M_1 - \Omega_1)$.
Note that the integration over $r$ is
formally divergent for the cases $\kappa = -1, 0$, corresponding to
open universes. To avoid possible technical complications associated
with such divergences, we consider for the moment
the case when the background FRW
 spacetime is spatially bounded, though the point we will be making
below is independent of the value of $\kappa$.
In this case, the integral over $r$
is convergent and gives (remembering that due to the spherical excluded
region, $a \leq r \leq 1$)
\begin{equation}
\int_a^1 \frac{dr\, r^2}{\sqrt{1 - r^2}} =
 \frac{\pi}{4} + \frac{a}{2} \sqrt{1 - a^2}
 +\frac{1}{2} \sin^{-1}(a) \equiv f(a),
\end{equation}
so that
\begin{equation}
S^{\kappa =1}_{background} =
 3\int d\tau \, f(a) \, \left( \frac{1 - a^2 +({\dot a}R)^2}{1 -
a^2} \right)^{1/2}\, (R^2 R'' + R{R'}^2 + R),
\end{equation}
employing the time calibration jacobian in (14). With (15) and (19), we
can now easily extract the
 wormhole lagrangian from
\begin{equation}
S_{wormhole} = \int d\tau \, L_{wormhole}(a,{\dot a}\,; R, R'),
\end{equation}
and calculate the momentum conjugate to $a$,
$\Pi = \partial L/{\partial {\dot a}}$ :
\begin{equation}
\Pi^{\kappa =1} = 2aR^2 \sinh^{-1} \left(\frac{{\dot a}R}{\sqrt{1 -
a^2}}\right)
 - \frac{a^2 R^2 R'}{\sqrt{1 - a^2}}
 + \frac{3{\dot a}R^2 f(a)}{\sqrt{1 - a^2 + ({\dot a}R)^2}}
 F^{\kappa = 1}(R,R',R''),
\end{equation}
where
\begin{equation}
F^{\kappa}(R,R',R'') = R^2 R'' + R{R'}^2 + \kappa R.
\end{equation}
As it stands, the transcendental relation (21) in
general cannot be inverted in closed
form in order to yield ${\dot a}$ in terms of $\Pi$ and $a$.
The `culprit' of this obstruction is the jacobian factor
$(dt/d{\tau})$, which depends explicitely on the throat velocity $\dot
a$ for all values of $\kappa$.
However, provided there exist backgrounds for which
 $F^{\kappa} = 0$, then inversion of this relation in closed
form is possible, and we
can solve for the throat velocity in terms of the throat momentum
(valid for all $\kappa$):
\begin{equation}
{\dot a} = \frac{\sqrt{1 - \kappa a^2}}{R}\,
 \sinh \left( \frac{\Pi}{2aR^2} + \frac{aR'}{2\sqrt{1 - \kappa a^2}}
\right),
\end{equation}
where the canonical momentum $\Pi$ is that of the wormhole's throat.
Fortunately, there are several physically interesting cases of
background
FRW cosmologies for which $F^{\kappa}$ vanishes identically. Indeed,
they
are (i) Minkowski space: $\kappa = 0, R = 1$ \cite{Visser90a}, (ii) radiation
dominated expansion: $\kappa = 0, R(t) = t^{1/2}$, and
(iii) DeSitter space: $\kappa = 0, R(t) = e^{\sqrt{|\Lambda|/3}\, t}$,
where $\Lambda$ is a cosmological constant. For this latter case, one
of course adds a cosmological constant term to the action in (8).
We shall consider wormhole
dynamics with $F^{\kappa = 0} = 0$ for the remainder of this paper. We
should point out that cases (ii) and (iii) are, of course, of particular
relevance
when treating quantum wormholes in the early universe. Moreover, as
the curvature $\kappa$ is negligible during the early
stages of expansion, taking $\kappa = 0$ is a good approximation
during this epoch.

{}From the lagrangian in (20)
the classical Wheeler-DeWitt hamiltonian for the wormhole is obtained by the
standard Legendre transform :
\begin{eqnarray}
H(\Pi,a; R,R') & = &  \Pi {\dot a} - L(a,{\dot a},R,R') \\
         & = & -2a^2 RR' \sqrt{1 + ({\dot a}R)^2}\, \sinh^{-1}({\dot
a}R) \nonumber \\
   &  & \mbox{ } + 2{\dot a}R'(a^2R^2) + 2aR\sqrt{1 + ({\dot a}R)^2} \nonumber
\\
  & = & -2a^2RR' \cosh \left( \frac{\Pi}{2aR^2} + \frac{aR'}{2} \right)
\nonumber
 \left( \frac{\Pi}{2aR^2} + \frac{aR'}{2} \right) \\ \nonumber
 & & \mbox{ } +  2a^2 RR' \sinh \left(\frac{\Pi}{2aR^2} +
 \frac{aR'}{2}\right)\\
 \nonumber
   & &\mbox{ } + 2aR \cosh \left(\frac{\Pi}{2aR^2} + \frac{aR'}{2} \right),
\nonumber
\end{eqnarray}
where in the final equality we used (23) to eliminate $\dot a$ in terms
of $\Pi$.
As is to be expected, the classical dynamics of the wormhole follows
by
setting $H(\Pi,a; R,R') = 0$, the vanishing of the hamiltonian being a
consequence of the reparametrization invariance of the gravitational
action. Note that this hamiltonian depends explicitely on external
cosmic time through the scale factor $R$ and its first derivative
$R'$.
Moreover, the canonical momentum of the throat (21) is {\it shifted} by an
amount directly proportional to time rate of change of the scale
factor.
The one degree of freedom in this (classical) minisuperspace is the throat
radius function $a$.

\section{Quantum Dynamics of the Wormhole}

 As we are interested in the quantum mechanical features of
Minkowski-signature wormholes, let us proceed directly to quantize
canonically the class of wormholes constructed above. Invoking the
prescription
\begin{equation}
\Pi \rightarrow -i\hbar \frac{\partial}{\partial a},
\end{equation}
promotes the classical constraint $H = 0$ to a differential equation
\begin{equation}
H \left (-i\hbar \frac{\partial}{\partial a}, a ;R,R' \right) \Psi (a,t) = 0,
\end{equation}
where $\Psi(a,t)$ is the time-dependent wavefunction of the wormhole.
As pointed out above,
 the throat momentum $\Pi$ is shifted relative to the $\sinh^{-1}$
factor by an amount proportional to the time rate of change of the
background spacetime, $R'$. This term represents the coupling between
the wormhole throat and the background. This shift results in a {\it
complex}
Wheeler-DeWitt hamiltonian operator
\begin{equation}
H = H_r + i H_i
\end{equation}
where
\begin{eqnarray}
 H_r =  [ -(2a^2RR')x\cosh(x) &+& (2a^2RR')\sinh(x) +
          (2aR)\cosh(x)  ]
 \cos({\bf y}) \\
  &+&  (2a^2RR')\sinh(x)\, {\bf y}\sin({\bf y}), \nonumber
\end{eqnarray}
is the real part and
\begin{eqnarray}
H_i =  [ (2a^2RR')x\sinh(x) &-& (2a^2RR')\cosh(x)
 - (2aR)\sinh(x)  ]
 \sin({\bf y}) \\
  &+& (2a^2RR')\cosh(x)\, {\bf y}\cos({\bf y}) \nonumber
\end{eqnarray}
is the imaginary part. Here, $x = aR'/2$ and the operator ${\bf
 y}=\frac{L^2_P}{2aR^2}
\frac{\partial}{\partial a}$, with $L_P = {\hbar}^{1/2}$
the Planck length, and we have
adopted the factor ordering prescription such that ${\bf y}$ always
stands to the right of $x$.
In the absolute static limit (e.g., case (i)), $R' = 0$
 so $x = 0$, and we obtain (after setting $R=1$, without loss of generality)
\begin{eqnarray}
H_r &=& 2a\cos \left( \frac{L^2_P}{2a} \frac{\partial}{\partial a} \right)\\
H_i &=& 0,
\end{eqnarray}
that is, the hamiltonian is purely real.
Solutions of the Wheeler-DeWitt equation with the hamiltonian in
(30,31) provide the starting point for a time-independent
quantum analysis of wormholes
in a Minkowski background \cite{Visser90a}.

Although we have not been able to
find exact solutions of (26) for $R' \neq
0$, we can obtain approximate solutions for an adiabatic expansion of the
hamiltonian. To this end, we expand $H_r + iH_i$ to order $R'$, and put
$H = H^{(0)}_r +iR'\, H^{(1)}_i$ where
\begin{eqnarray}
H^{(0)}_r & = & 2aR \cos ({\bf y}), \\
H^{(1)}_i & = & -(3a^2R) \sin({\bf y}) + (2a^2R)\, {\bf y}\cos({\bf y}).
\end{eqnarray}
We solve for the wormhole wavefunction to $O(R')$ in perturbation
theory:
\begin{equation}
\Psi(a,t) = \Psi^{(0)}(a,t) + R'\, \Psi^{(1)}(a,t),
\end{equation}
where $\Psi^{(0)}$ is the exact zeroth-order solution which satisfies
$H^{(0)}_r \Psi^{(0)}(a,t) = 0$. To this order, the Wheeler-DeWitt
equation
implies a relation between the zeroth and first order parts of the
wavefunction:
\begin{equation}
H^{(0)}_r\, \Psi^{(1)}(a,t) = -i\, H^{(1)}_i\, \Psi^{(0)}(a,t).
\end{equation}
We make use of the fact that
there is an infinite set of exact zeroth-order solutions
\begin{equation}
\Psi^{(0)}_m(a,t) = C_m\, e^{-(m+ \frac{1}{2})\pi a^2 R^2/{L^2_P}},
\end{equation}
which is an obvious
generalization of the wavefunction employed in \cite{Visser90a} .
The integer index $m = 0,1,2, \cdots $, negative values being discarded as the
wavefunction is not normalizable for $m < 0$. Since
\begin{equation}
{\bf y}\, \Psi^{(0)}_m (a,t) = -(m + \frac{1}{2}) \pi \, \Psi^{(0)}_m (a,t),
\end{equation}
inserting (36) into (35) leads to
\begin{equation}
\cos({\bf y})\, \Psi^{(1)}_m (a,t) = (-)^{m+1} \frac{3ia}{2} \,
 \Psi^{(0)}_m (a,t).
\end{equation}
To solve this equation, we assume a product ansatz
\begin{equation}
\Psi^{(1)}_m (a,t) = g(a,t)\, \Psi^{(0)}_m (a,t),
\end{equation}
which when substituted into (38) and employing the identity \cite{foot2}
\begin{equation}
\cos({\bf y}) g(a)f(a) = [\cos({\bf y})g(a)][\cos({\bf y})f(a)]
 -[\sin({\bf y})g(a)][\sin({\bf y})f(a)],
\end{equation}
leads to the equation
\begin{equation}
\sin({\bf y})\, g(a,t) = -\frac{3i}{2}\,a.
\end{equation}
This nonlocal inhomogeneous differential equation can be solved using
Green function techniques.
The details of the calculation of the Green function for the operator
$\sin({\bf y})$
 are relegated to the Appendix. The result is that
\begin{equation}
g(u) =  \frac{3L_P}{2iR}\,\int_0^{\infty} G(u - u')\, \sqrt{u'} \,du'
 + \sum A_m e^{-m\pi u},
\end{equation}
where $u = a^2R^2/{L_P}^2 \geq 0$ and the Green function $G(u-u')$ is
calculated explicitely in (A10,A11).

To proceed, we take the $A_m = 0$ (i.e., vanishing homogeneous
solution) and note that the Green function (A10,11) has the form
of a ``kink'' or topological soliton with center at $u > 0$:
\begin{equation}
\lim_{u' \rightarrow \pm \infty} G(u - u') =
  \left\{ \begin{array}{ll}
   -\sqrt{\frac{\pi}{2}}, & \mbox{for $u' > u$} \\
   +\sqrt{\frac{\pi}{2}}, & \mbox{for $u' < u$}
   \end{array}
   \right. .
\end{equation}
While (42) is exact, the integral cannot be reduced to known elementary
functions. Nevertheless, a reasonable approximation can be had by
taking
\begin{equation}
G(u - u') \sim \sqrt{\frac{\pi}{2}}
 \left( \Theta(u - u') - \Theta(u' - u) \right),
\end{equation}
which does incorporate the essential kink structure and asymptotic behavior
of $G$. Doing so, we obtain
\begin{equation}
g(a,t) = -i\sqrt{2\pi}\, a^3\, (R(t)/L_P)^2,
\end{equation}
after discarding an infinite constant \cite{foot3}.

 The wormhole wavefunction, to the order we are working, is therefore given by
\begin{equation}
\Psi_{mn}(a,t) = \Psi^{(0)}_{mn}(a,t) \left( 1 -i\sqrt{2\pi} a^3 R' (R/L_P)^2
\right),
\end{equation}
where
\begin{equation}
\Psi_{mn}^{(0)}(a,t) = C_{mn}\,
 \left[ e^{-(m+\frac{1}{2})\pi (\frac {a R}{L_p})^2} -
 e^{-(n+\frac{1}{2})\pi (\frac{a R}{L_P})^2 } \right],
\end{equation}
and is manifestly complex, the imaginary part being induced by the
time
dependence of the background spacetime. The leading order factor is
simply
a linear combination of the exact zeroth-order solutions so chosen that
$\Psi^{(0)}_{mn}(0) = 0$, $(m \neq n)$ the condition
required for hermiticity of
$H^{(0)}_r$ \cite{Visser90a}.

With the FRW wormhole wavefunction in hand, we proceed to calculate
the
normalization constant $C_{mn}$ and the expectation value of the
wormhole throat radius.
{}From
\begin{equation}
\int_0^{\infty} da\, |\Psi_{mn}(a,t)|^2 = 1,
\end{equation}
we find after an elementary integration that
\begin{eqnarray}
 C^{-2}_{mn} &=&  \frac{L_P}{2R} \left(\frac{1}{\sqrt{2m +1}} +
 \frac{1}{\sqrt{2n+1}} - \frac{2}{\sqrt{n+m+1}} \right)  \\
 &+&  \frac{15L_P^3}{8\pi^2}\frac{{R'}^2}{R^3}
 \left( (2m+1)^{-7/2} + (2n+1)^{-7/2} -2(n+m+1)^{-7/2} \right) . \nonumber
\end{eqnarray}

In a similar fashion, we find for the matrix element of the throat radius
\begin{eqnarray}
<\Psi_{mn}|aR|\Psi_{mn}> \,= \,
<aR>_{mn} &=& R<a>_{mn} \,= \, R\int_0^{\infty} da \,a\, |\Psi_{mn}(a,t)|^2 \\
 &=& \frac{L_P}{\pi} \frac{[\frac{1}{2m+1} +\frac{1}{2n+1}
-\frac{2}{n+m+1}]}
{[\frac{1}{\sqrt{2m+1}}
+\frac{1}{\sqrt{2n+1}}-\frac{2}{\sqrt{n+m+1}}]}  \nonumber \\
&\times& \{ 1 + \frac{L_P^2}{\pi^2}(\frac{R'}{R})^2\,f_{(m,n)} \} \nonumber
\end{eqnarray}
where the numerical factor $f_{(m,n)}$ is defined by
\begin{equation}
f_{(m,n)}=12\,\frac{[\frac{1}{(2m+1)^4}+\frac{1}{(2n+1)^4}-\frac{2}{(n+m+1)^4}]}
{\frac{1}{(2m+1)}+\frac{1}{(2n+1)}-\frac{2}{(n+m+1)}}
-\frac{15}{4}\frac{\frac{1}{(2m+1)^{7/2}}+\frac{1}{(2n+1)^{7/2}}
-\frac{2}{(n+m+1)^{7/2}}}{\frac{1}{\sqrt{2m+1}}+\frac{1}{\sqrt{2n+1}}
-\frac{2}{\sqrt{n+m+1}} }.
\end{equation}
Thus, to lowest order, the average value of the throat radius is of
the order the Planck length with corrections coming from the
explicit time dependence of the background spacetime.
{}From (50), we can see that for static backgrounds, we recover
the result that the wormhole is quantum mechanically stabilized
against
collapse, with an average radius $= L_P$ \cite{Visser90a}.
However, when the background
depends
on time, the wavefunction becomes manifestly time dependent and the
wormhole throat radius need no longer be stationary. Indeed, already
at
lowest order, we find that the radius is modified from its static
value.
Thus, for radiation dominated expansion (RDE), $(\frac{R'}{R})^2 =
\frac{1}{4t^2}$,
while for a deSitter cosmology, $(\frac{R'}{R})^2 = |\Lambda|/3$.
As $f_{(m,n)} > 0$, we see from (50) that the expansion augments the
magnitude of the matrix element.
It is noteworthy that for the radiation dominated case, the
perturbation to the throat radius expectation value
vanishes as $t \rightarrow \infty$ so that the wormhole
throat {\it relaxes} from initially larger values to assume
its static value. For the deSitter
example, the first order perturbation is
actually time independent; in summary:
\begin{eqnarray}
<aR>_t &\rightarrow& <aR>_{static} \,\,\approx \,L_P, \qquad {\rm RDE},\\
<aR>_t &>&   <aR>_{static} \, \qquad {\rm deSitter}.
\end{eqnarray}
These calculations indicate that the expectation value for the wormhole
radius is larger than its static value when the background is
expanding, indicating that the throat gets partially dragged open with the
expansion. This is perhaps not too surprising as
the wormhole is gravitationally coupled to the
background. However, the {\it extent}
of the dragging cannot be known {\it a priori}, and we clearly see that it
depends
crucially on the form of the background scale factor.

\section{Comoving Wormholes}

When matter is added to the throat, even approximate solutions of the
Wheeler-DeWitt equation are difficult to obtain. The variant of the
WKB approximation scheme described
in \cite{foot2}, designed to handle
non-quadratic hamiltonians, fails to be of use for our
hamiltonian because the classical energy relation
\begin{equation}
H \left( \Pi,a;R,R' \right) = E
\end{equation}
with $H$ as given in (24),
cannot be inverted in closed form so as to express
$\Pi = \Pi (E,a; R,R')$, the latter relation needed for calculating the
WKB wavefunction. Nevertheless, we can make an exact statement, at
least at the classical level, concerning the state of motion
and type of matter needed such that
the wormhole throat is comoving, that is, coupled to the Hubble flow.

The most general stress-energy tensor which gives rise to two
identical FRW spaces connected by a $\delta$-layer wormhole is
\begin{equation}
T^{\mu}_{\nu}(x) = S^{\mu}_{\nu}(x) \, \delta (\eta) +
T^{(1)\mu}_{\nu} \, \Theta (\eta) + T^{(2)\mu}_{\nu} \, \Theta
(-\eta),
\end{equation}
where $T^{(1)\mu}_{\nu} = T^{(2)\mu}_{\nu}$ is one of the standard
perfect fluid source terms leading to (4) and
\begin{equation}
S^{\mu}_{\nu} \equiv \lim_{\epsilon \rightarrow 0} \,
 \int^{\epsilon}_{-\epsilon} \, d{\eta} \,\, T^{\mu}_{\nu} (x),
\end{equation}
is the surface stress energy on the wormhole throat. In components
\begin{equation}
S^i_j = {\rm diag} (-\sigma, P, P )
\end{equation}
where $\sigma$, $P$ denote the surface energy and pressure densities.
The throat matter action and lagrangian are
\begin{equation}
S_{matter} = -\int d^3x \, \sqrt{-h} \, \sigma = -4\pi \int d\tau \,
a^2R^2 \, \sigma = \int d\tau \, L_{matter}.
\end{equation}
The Wheeler-DeWitt hamiltonian for the combined system of wormhole
plus matter $H_{total} = H_{throat} + H_{matter}$, where
\begin{equation}
H_{matter} = -L_{matter} = 4\pi a^2 R^2 \, \sigma.
\end{equation}

In order that the wormhole be comoving with the background expansion,
${\dot a} = 0$, since the radius $l(t) = a R(t)$ is then simply
proportional to the scale factor. This is also reflected in the fact
that $dt/{d\tau} = 1$ if and only if $a$ is constant, i.e., throat
proper time is identical to cosmic time when there is no relative motion
between the throat and background.
For $\dot a = 0$ the classical hamiltonian constraint, obtained after
summing (24) and (59), reduces to
\begin{equation}
2aR + 4\pi a^2 R^2 \sigma = 0,
\end{equation}
that is, the surface energy density on the throat must be
\begin{equation}
\sigma = -\frac{1}{2\pi a R} < 0
\end{equation}
which is negative definite and scales as $R^{-1}$ in time. According to
this, the wormhole mouth will expand just as rapidly as the Hubble
flow and for all scale factors. However, in order for this motion to be
realized, the matter on the throat must be ``adjusted'' to have the
surface energy density specified in (61) as well as obey the equation of
state $P = -\frac{\sigma}{2}$ \cite{Hoch93}. We suspect this will also
hold at the quantum level, that is, a quantum comoving wormhole, one
for which $<aR> = {\rm constant} \times R$, will require nontrivial
stress-energy located at the throat. We can turn this argument around
and conclude that in general, the wormhole will not be comoving, but
will exhibit motion relative to the background expansion.

This result, that a classical comoving wormhole entails weak energy
condition (WEC) violating matter should be compared to the results of Roman
\cite{Roman93}, who analyzed a class of Morris-Thorne type metrics
representing Lorentzian wormholes embedded in a deSitter inflationary
background. Those classical wormholes were shown to inflate in step
with the background and require a non trivial stress energy tensor
violating the WEC.

\section{Discussion}

In this paper we have investigated a simple model of
Minkowski-signature wormholes attached to FRW spacetimes. The quantum
gravitational dynamics of these wormholes is reduced to the quantum
mechanics of a single variable, namely, the throat radius, by solving
the Wheeler-DeWitt equation over minisuperspace. When quantizing
wormholes in {\it static} backgrounds via the Wheeler-DeWitt
prescription \cite{Visser90a, Kim92}, one comes up against the problem
of time. That is, the resulting wavefunction is time-independent,
despite the fact that the wormhole may well evolve classically. This
makes it awkward to discuss the quantum stability since the
calculation of {\it transitions} between allowed quantum states of the
wormhole lies beyond the scope of the formalism. However, for
time-dependent backgrounds one can treat the evolution of quantum
wormholes in an obvious and natural way. For the case of FRW
backgrounds, the temporal parameter is provided by the scale factor.
Moreover, FRW cosmologies are the relevant spacetimes in which to
analyze early universe wormholes.

We have calculated the average wormhole radius for both radiation
dominated and deSitter inflationary scale factors. It is noteworthy
that even in the absence of matter at the throat, the
wormholes are stabilized quantum mechanically against both collapse
and blowing-up. For radiation domination, the wormhole radius
decreases monotonically from intially large values to attain a minimum
of order the Planck length. For the inflationary case, the
lowest-order calculation reveals the radius to be held fixed at a
value always larger than the Planck length, by an amount proportional
to the absolute value of the cosmological constant (vacuum energy
density). These two cases indicate that the wormhole throat gets
dragged open by the background expansion, although there is relative
motion, or ``slippage'', between the wormhole and the ambient Hubble
flow. This result is encouraging as it weighs in favor of the ideas
set forth in \cite{Hoch93}.

A limitation of our calculation is the fact that the wormholes treated
here have zero-length handles. Of course, we by no means claim that
all (Planck-scale) wormholes would be of this form, but we believe
that the qualitative aspects of their evolution should not be overly
sensitive to this trait. Nevertheless, it would be interesting to
check this by quantizing finite handle wormholes as described, for
example, by Morris-Thorne type metrics.

Perhaps a more severe limitation of our calculation is due to the fact
that we have not employed the full superspace description for the
wormhole dynamics. The truncation from an infinite number of degrees
of freedom down to a single degree of freedom may appear drastic, but
it is the only viable option if we wish to be able to make a definite
calculation. Working in the full superspace is obviously preferable,
but is currently an intractable problem.

Finally, when formulating a time-dependent quantum mechanics of
wormholes via the Wheeler-DeWitt approach, as we have done here, we find
the hamiltonian operator picks up many new terms not present in
the static background limit, e.g., compare (28,29) to (30,31). These
additional terms are functions of the expansion velocity $R'$ and
complicate the task of solving for the wavefunction. Nevertheless, we
have been able to calculate the wavefunction perturbatively employing
$R'$ as a small parameter. Naturally, one would like to get at
information lying beyond the restraints of perturbation theory. To
this end, we would like to remark that the wave equation (26) may lend
itself to approximate, but non-perturbative, solutions based on methods
borrowed from the calculus of finite differences \cite{privcom}. This
follows from the observation that the nonlocal operators appearing
there can be expressed in terms of finite translations acting on
minisuperspace. These points will be explored elsewhere.

\vspace{1cm}
\noindent
{\bf Acknowledgements}
The author is grateful to Jos\'e Luis Rosales for many extensive and
stimulating discussions. I would also like to thank Joaquin Retamosa
for valuable comments.

\appendix
\section{Appendix}
\renewcommand{\thesection}{\Alph{section}}

We wish to obtain an explicit representation for $G$, defined as the
inverse, over $C^{\infty}$ function space, of the nonlocal differential
operator $\sin(\frac{d}{du})$, that is,
$$
\sin \left( \frac{d}{du} \right) G(u,u') = \delta(u - u'),
\eqno(A1)$$
where
$$
\sin \left( \frac{d}{du} \right) = \frac{d}{du} - \frac{1}{3!}
\frac{d^3}{du^3} + \frac{1}{5!}\frac{d^5}{du^5} - \cdots
\eqno(A2)$$
As the operator (A2) is linear, it proves useful to
introduce the Fourier transform $\tilde G$ of $G$ via
$$
G(u,u') = \int_{-\infty}^{\infty} \frac{dp}{\sqrt{2\pi}}\, {\tilde G}(p)
e^{-ip(u - u')},
\eqno(A3)$$
whereby using the identity
$$
\sin \left( \frac{d}{du} \right) e^{-ipu} =\sin (-ip) e^{-ipu},
\eqno(A4)$$
we have that the transform of the Green function is
$$
{\tilde G}(p) = \frac{i}{\sinh(p)}.
\eqno(A5)$$
The problem of computing $G$ thus reduces to carrying out the integration in
(A3) with ${\tilde G}$ as given in (A5). There are two cases to consider:

\subsection{case (1): $\Delta = (u - u') > 0$}

We go to the complex $p$-plane and consider the following contour
built up from closed intervals and semicircular arcs centered about the origin:
$\Gamma_{R,\delta}= [-R,-\delta] + \gamma_{\delta} + [\delta,R]
+ \gamma_R$, where $\gamma_{\delta}$ is a semicircle of radius
$\pi > \delta > 0$, and $\gamma_R$ a semicircle of radius $R$
with $(M+1)\pi > R > M\pi$ $(M >> 1)$, both located in the lower half-plane,
as
indicated in Fig. A. The inequalites are imposed so that the contour
avoids hitting any of the poles.
Since $\sinh(x+iy)=\sinh(x)\cos(y) +
i\cosh(x)\sin(y)$, the integrand in (A3) has simple poles located on the
imaginary axis: $z = \pm im\pi$ for $m = 0, 1, 2, \cdots$.
Thus, by the residue calculus \cite{Marsden73}, we can calculate
the integral taken around the complete contour in the clockwise sense
$$
\oint_{\Gamma_{R,\delta}} \frac{dz}{\sqrt{2\pi}}
 \, \frac{ie^{-iz\Delta}}{\sinh(z)}
  = \sqrt{2\pi}  \sum^M_{m=1}
(-e^{-\pi \Delta})^m.
\eqno(A6)$$
The Fourier integral representation for $G$ (A3) results after taking the two
 limits $\delta \rightarrow 0$ and
$R \rightarrow \infty$. Denoting by $I_R$ the contribution to
(A6) coming from the large semicircular path $\gamma_R$, it is
straightforward to show that
$$
0 \leq |I_R| \leq R\int_0^{\pi} \frac{d\theta}{\sqrt{2\pi}}\,
\frac{e^{-\Delta R \sin \theta}}{|\cosh^2(R\cos \theta) - \cos^2(R
\sin \theta) |^{1/2}} \longrightarrow 0,
\eqno(A7)$$
as $R \rightarrow \infty$ for every $\theta \in  [0,\pi]$.
On the other hand, the contribution $I_{\delta}$ to (A6) coming from
the small semicircle $\gamma_{\delta}$ centered about the origin gives
$$
\lim_{\delta \rightarrow 0} I_{\delta} = -\sqrt{\frac{\pi}{2}}.
\eqno(A8)$$
Putting together (A3) and (A6-8), we have that
$$
\sqrt{2\pi} \sum^{\infty}_{m=1} (-e^{-\pi \Delta})^m =
\int^{\infty}_{-\infty} \frac{dp}{\sqrt{2\pi}}
 \frac{ie^{-ip\Delta}}{\sinh(p)} + \lim_{\delta \rightarrow
0}\,I_{\delta}
+ \lim_{R \rightarrow \infty} \,I_R,
\eqno(A9)$$
so that
$$
 G(\Delta) = \sqrt{\frac{\pi}{2}} - \frac{\sqrt{2\pi}\, e^{-\pi \Delta}}{1
 + e^{-\pi \Delta}},
\eqno(A10)$$
where in the limit $M \rightarrow \infty$ , $R \rightarrow \infty$ and
$\delta \rightarrow 0$ the contour $\Gamma_{R,\delta}$ maps to the
real line joined to an infinite semicircle in the lower half-plane.

\subsection{case (2): $\Delta = (u - u') < 0$}

   This time, we close the contour in the upper half-plane and in the
counterclockwise sense, i.e.,
$\Gamma_{R,\delta} = [-R,-\delta] + \gamma_{\delta} + [\delta, R]
+ \gamma_{R}$, where now $\gamma_{\delta}$ and $\gamma_R$ are semicircular
arcs in the upper half-plane. The steps and limits involved in
computing
$G$ are entirely analogous to those of the previous case, with the
final result that
$$
 G(\Delta) = -\sqrt{\frac{\pi}{2}} + \frac{\sqrt{2\pi}\, e^{\pi
\Delta}}{1 + e^{\pi \Delta}}.
\eqno(A11)$$

The Green function (A10,11) is bounded for $|\Delta| \rightarrow \infty$ and
$G(0) = 0$. From (A1), the general solution
of $sin(\frac{d}{du})\, g(u) = f(u)$ is therefore
$$
g(u) = g_{homogeneous}(u) + \int du' \, G(u-u')\,f(u'),
\eqno(A12)$$
where the homogeneous part of the solution is given by
$$
g_{homogeneous} = \sum A_m \, e^{-m\pi u}.
\eqno(A13)$$

\vfill\eject

\vfill\eject
\noindent
{\bf Figure Caption}
Fig. A.  The contour $\Gamma_{R,\delta}$ for case (1).

\begin{thebibliography}{99}
\bibitem{foot1} It is not necessary to take $R_1 = R_2$, or $\kappa_1
= \kappa_2$, but doing so simplifies the analysis.
\bibitem{foot3} A suitable cut-off for the integral in (42) is
provided by the horizon radius. Nevertheless, as our final results are
qualitatively unaffected by the inclusion of an additive constant in
(45), we omit this from our consideration.
\end{thebibliography}
\end{document}